\documentclass[12]{iaus}
\usepackage{graphicx}
\newcommand{\fat}[1]{{\bf #1}}
\newcommand{\half}{{\scriptscriptstyle{\frac12}}}

\title[Black holes] 
{Dancing with black holes}

\author[Sverre Aarseth]   
{Sverre J. Aarseth}

\affiliation{Institute of Astronomy, Madingley Road, Cambridge CB3 0HA, UK
\break email: sverre@ast.cam.ac.uk \\[\affilskip] }

\pubyear{2008}
\volume{246}  
\pagerange{001--10}
\date{?? and in revised form ??}
\jname{Proceedings Title IAU Symposium}
\editors{E. Vesperini, M. Giersz \& A. Sills, eds.}
\begin{document}

\maketitle

\begin{abstract}
We describe efforts over the last six years to
implement regularization methods suitable for studying one or more
interacting black holes by direct N-body simulations.
Three different methods have been adapted to large-N systems:
(i) Time-Transformed Leapfrog, (ii) Wheel-Spoke, and (iii) Algorithmic
Regularization.
These methods have been tried out with some success on GRAPE-type computers.
Special emphasis has also been devoted to including post-Newtonian terms,
with application to moderately massive black holes in stellar clusters.
Some examples of simulations leading to coalescence by gravitational
radiation will be presented to illustrate the practical usefulness of
such methods.
\end{abstract}

\firstsection 
\section{The chain concept}

In the study of strong gravitational interactions, utilization of the chain
data structure can be very beneficial.
Over many years, the original chain regularization method
(Mikkola \& Aarseth 1993) has proved to be effective in star cluster
simulations containing binaries.
As we shall see in the following, it is also a useful tool in connection
with time transformations which do not employ the usual coordinates.
By introducing one or more dominant masses these advantages become more
apparent.
Such problems fall naturally into three classes according to the number
of massive objects and each class requires special attention.
At the simplest level we have the case of one central massive body which
dominates the motion of other members within a certain distance.
The role of the reference body is readily seen in the case of three
interacting particles which can be studied by three-body regularization
(Aarseth \& Zare 1974).
This idea was extended to an arbitrary membership (Zare 1974).
However, a natural application was lacking until the problem of black
holes (BHs) became a challenge for simulators in recent years.
The aptly named wheel-spoke regularization (Aarseth 2003a) has now been
adapted to study compact subsystems containing a single massive object.

Historically speaking, a special method for a BH binary was implemented
in an $N$-body code first.
Here the main idea is based on a time-transformed leapfrog scheme (TTL)
suitable for dealing with large mass ratios (Mikkola \& Aarseth 2002).
Remarkably, this method yields machine precision for unperturbed two-body
motion.
Although regularized chain coordinates are not employed directly, the
accuracy is improved by using relative quantities with respect to the
nearest massive body.

Alternative methods may be needed for problems involving more than two
massive objects.
The recent algorithmic regularization (Mikkola \& Tanikawa 1999,
Preto \& Tremaine 1999, Mikkola \& Merritt 2006) appears to be a promising
way of studying such systems.
Indeed, the masses only play a kinematical role in one formulation,
suggesting that it may be applicable to systems with large mass ratios.
However, it should be emphasized that for practical reasons any of the above
methods are of necessity limited to relatively small particle numbers, or
in other words, compact subsystems.

\section{Post-Newtonian formulation}\label{sec:PN}

Relativistic effects in stellar systems are usually associated with high
densities or very massive objects.
However, quite large values of the eccentricity may also occur for
certain types of initial conditions (Aarseth 2003b, Berczik et al. 2006,
Iwasawa, Funato \& Makino 2006) such that the orbital shrinkage by
gravitational wave emission becomes significant.
This stage is characterized by velocities reaching an appreciable fraction
of the speed of light, $c$.
The original expressions for modelling post-Newtonian effects (Soffel 1989)
were subsequently replaced by an equivalent scheme which facilitates the
evaluation of consistent two-body elements
(Blanchet \& Iyer 2003, Mora \& Will 2004).
This development enables the equation of motion to be written in the concise
form
\begin{equation}
\frac {d^2 {\bf r}}{d t^2} =
 \frac {M}{r^2} \left[(-1 + A) \frac {\bf r}{r} + B {\bf v} \right ] ,
\end{equation}
where the scaled quantities $A, B$ represent the post-Newtonian terms.
An expansion of increasing complexity up to $1/c^6$ is available for
implementation.

Given the perturbing force ${\bf F}_{\rm GR}$, the corresponding energy loss
can be obtained by integrating ${\bf F}_{\rm GR} \cdot {\bf v} dt$ in
regularized form.
Here the secular change is due to gravitational wave radiation represented
by the terms $A_{5/2}, B_{5/2}$, while the two first orders are connected
with precession.
The radiation time-scale in $N$-body units is (Peters 1964)
\begin{equation}
 t_{\rm GR}\,=\, 
\frac {5 c^5\, a^4} {64 m_i m_0(m_i+m_0)} \frac {(1 - e^2)^{7/2}} {g(e)},
\end{equation}
with $m_0$ usually the dominant mass and $g(e)$ a known function of the
eccentricity (about 4 for large values).
Even a typical hard binary with $a \simeq 1 \times 10^{-4}$ would only
yield $t_{\rm GR} \simeq 1000$ for $e = 0.999$ and the present model parameters
($c \simeq 15\,000$).
Rather extreme dynamical evolution is therefore required to reach the
relativistic regime.

\section{Multiple regularization schemes}

In the following we summarize the main ideas involved in the three multiple
regularization schemes outlined above.
Although the TTL method (Mikkola \& Aarseth 2002) does not deal specifically
with the removal of singularities, it allows arbitrarily close encounters
(including collisions) to be studied.
Here the time transformation $t' = 1/W$ is combined with leapfrog integration
of physical coordinates and velocities.
The key feature is to replace an explicit evaluation of the auxiliary
variable $W = \Omega(r)$ by the differential equation
\begin{equation}
\dot W = {\bf v} \cdot \frac {{\partial \Omega}} {{\partial {\bf r}}},
\end{equation}
which is usually a slowly varying quantity and integrated by
$W' = {\dot W}/ \Omega$.
The function $\Omega$ may be taken as the sum of all inverse two-body
distances when large mass ratios are present.
This leads to equations of motion for the relative coordinates and velocities,
${\bf r, v}$ of the form
\begin{equation}
{\bf r}_i' = \frac {{\bf v}_i'} {W}, \,\,\, \, \,
 {\bf v}_i' = \frac {{\bf F}_i} {\Omega} ,
\end{equation}
with ${\bf F}_i$ the usual $N$-body acceleration.

The solutions are formulated as a set of leapfrog equations, with the
quantity $W$ integrated over the time-step $h$ by
\begin{equation}
 W_1=W_0+h\frac{\fat v_0+\fat v_1}{2\Omega(\fat r_{\half})}\cdot
 \frac{\partial\Omega(\fat r_{\half})}{\partial\fat r_{\half}}.
\end{equation}

The addition of an external perturbation ${\bf f}_i$ gives rise to
the energy equation
\begin{equation}
E' = \sum m_i {\bf v}_i \cdot \frac {{\bf f}_i} {\Omega}
\end{equation}
which can be integrated in the same way as the other equations.
As in the post-Newtonian formulation, velocity-dependent terms can be
included by using the implicit mid-point method and solved by iteration.

\begin{figure}
\includegraphics{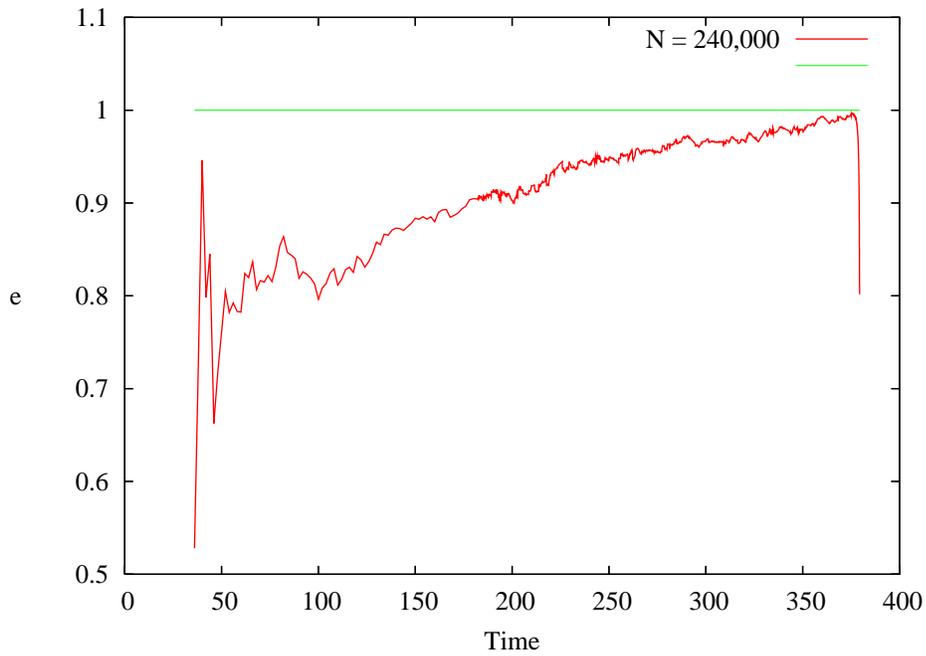}
\centering
  \caption{Eccentricity evolution of a massive BH binary for $N = 2.4 \times 10^5$.}
\label{fig:ecc4}
\end{figure}

In the case of one dominant body, it is natural to treat the system as a
wheel-spoke with all members connected to the central hub (Zare 1974).
This representation can now be seen as an alternative to the chain geometry.
Hence the introduction of regularized coordinates and momenta, ${\bf Q}$ and
${\bf P}$, as well as their transformations can be taken over directly from
chain regularization.
After multiplying by the time transformation function $g({\bf q, p})$
the resulting Hamiltonian can be written as
\begin{equation}
\Gamma^{\ast} = g({\bf q, p}) \left[ H({\bf Q}_i, {\bf P}_i) - E \right ] .
\label{Gamma}
\end{equation}
Here the Hamiltonian function itself should be equal to the energy, $E$,
along the solution path, with deviations due to numerical errors.
The choice of time transformation as the inverse Lagrangian (in scaled
physical units) has proved highly effective.

In contrast to standard two-body regularization where $g = r$
and the singularities are removed explicitly, the equations of motion
derived from (\ref{Gamma}) must be differentiated term by term without
employing the simplifying condition $\Gamma^{\ast} = 0$.
Again the effect of perturbers in changing the internal energy must be
taken into account and added to any post-Newtonian contributions.

The third method, algorithmic regularization (Mikkola \& Merritt 2006),
makes use of the implicit mid-point rule to achieve a time-symmetric
leapfrog scheme.
In this way, velocity-dependent terms of the post-Newtonian expansion
can be handled by the extrapolation method.
This elegant algorithm produces exact solutions for unperturbed two-body
motion as well as accurate results in strongly interacting few-body systems.
An attractive feature is that the method works for arbitrary mass ratios.
The new formulation employs two equivalent time transformations for
coordinates and velocities, respectively, given by
\begin{equation}
t_q' = \frac {1} {\alpha T + B}, \, \, \, \, \,
t_{v}' = \frac {1} {\alpha U + \beta \Omega + \gamma} ,
\end{equation}
where $\alpha, \beta$ and $\gamma$ are dimensionless constants and
$T, U$ and $B$ are the kinetic, potential and positive binding energy.
As above, $\Omega$ can be taken to be the inverse sum of all separations.
Moreover, the slowly varying quantities $\Omega$ and $B$ are obtained by
integration.
These time transformations are qualitatively similar to the inverse
Lagrangian used in wheel-spoke regularization.
We also note that the chain data structure is utilized in order to prevent
loss of accuracy.

\begin{figure}
 \includegraphics{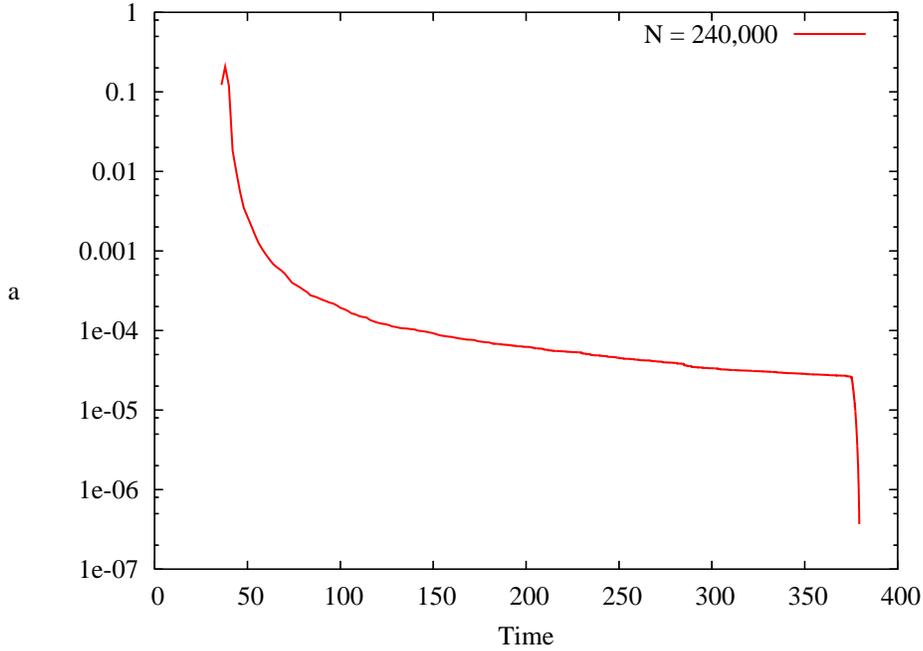}
  \caption{Semi-major axis of a massive BH binary for $N = 2.4 \times 10^5$.}
 \label{fig:semi4}
\end{figure}

\section{$N$-body implementations}

Each method requires a suitable subsystem to be present which is ideally of
a long-lived nature.
In the first instance, this usually occurs after the emergence of a hard
binary containing one massive object near the centre.
Other nearby members are then added to the subsystem using standard
selection criteria as for chain regularization.
Conversely, any particles moving away from the BH system are included with
the perturbers.
These procedures entail updating the internal energy consistently without
explicit evaluation of the dominant terms.
The membership change also requires initialization of force polynomials for
the new centre of mass and any ejected particle.
Since the simulations are made on a GRAPE-6A (or GRAPE-6 initially), this
procedure requires making differential force corrections on the host
computer for interactions between subsystem members and perturbers.
Finally, a check of total energy conservation is facilitated by separate
integration of the perturbations and relativistic contributions.

The special treatment for adapting the TTL method to a large-$N$ simulation
is analogous to that of chain regularization (Aarseth 2003a).
The equations of motion are integrated to high accuracy and include the
effect of perturbers selected in the usual way.
Conversely, the evaluation of the perturber forces take into account the
structure of the subsystem.
In order to study a BH binary with massive components, we choose two
identical spherical systems in an elliptic orbit with eccentricity
$e_{\rm orb} = 0.8$ and a massive object at the centre of a cusp-like
density distribution.
In such a case of two merging clusters, the massive binary formed on a short
time-scale (about 36 $N$-body units) and reached the hard binary stage
($a = 4 \times 10^{-4}$) at twice this time (see figure~\ref{fig:semi4}).

The algorithms for the wheel-spoke regularization have been described in
considerable detail elsewhere (Zare 1974, Aarseth 2003a, Aarseth 2007).
Here we emphasize that each subsystem interaction with the massive
reference body is treated as a standard KS regularization
(Kustaanheimo \& Stiefel 1965), while the other internal interactions
are subject to a small softening to smooth near-collisions of point-masses.
Again the first-order equations of motion are advanced by a high-order
integrator (Bulirsch \& Stoer 1966).
In view of the large mass ratio employed (around 300), only a few perturbers
are usually selected for 3--4 members of the subsystem in this preliminary
investigation.

\begin{figure}
 \includegraphics{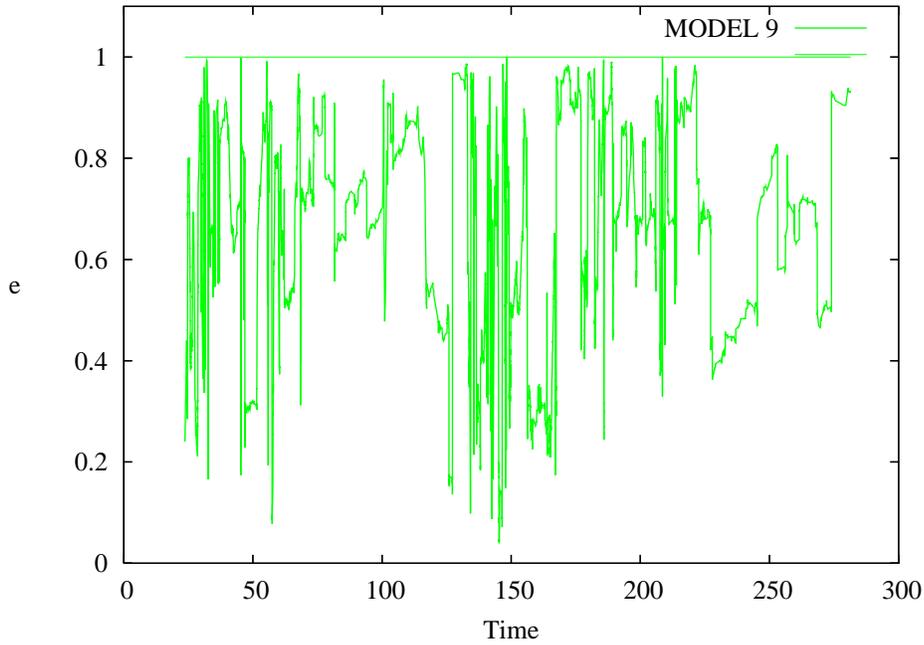}
  \caption{Eccentricity of the innermost binary, model M9.}
 \label{fig:ecc9}
\end{figure}

A sequential decision-making strategy has been implemented for including
relativistic effects, initially up to 2.5PN (Aarseth 2003b).
The main idea is to treat only the most dominant two-body motion.
Moreover, the relevant terms are included progressively according to the
value of $t_{\rm GR}$.
For convenience we choose $c = 3 \times 10^5/ V^{\ast}$, where $V^{\ast}$ is
the rms velocity in $\rm km \,s^{-1}$.
Hence for a specified total mass, relativistic effects can be examined via
different values of the half-mass radius and the equilibrium velocity
dispersion.
 
\begin{figure}
 \includegraphics{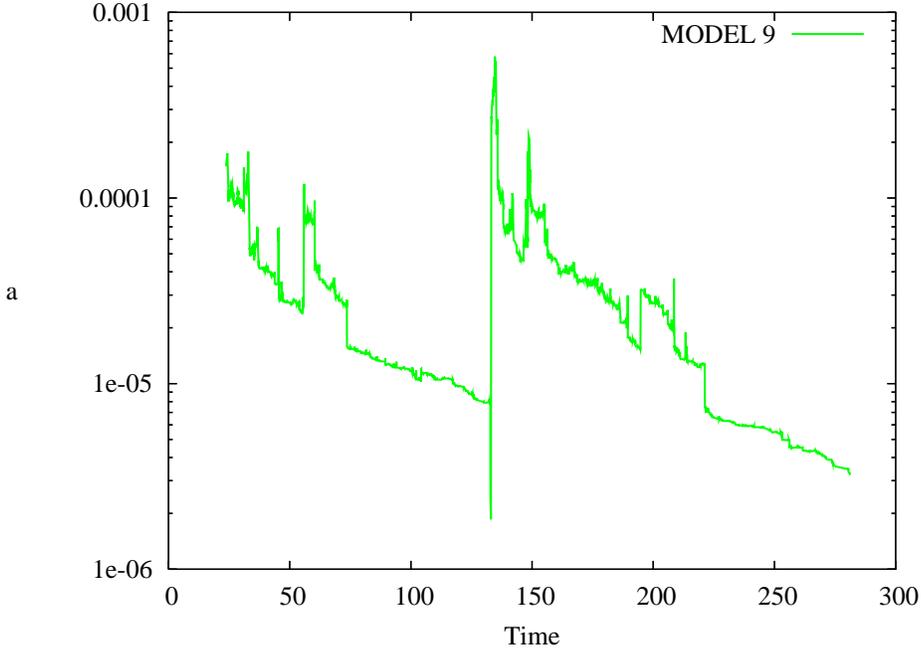}
  \caption{Semi-major axis of the innermost binary, model M9.
GR coalescence occurred at $t = 32,\,33,\,56,\,133,\,148,\,195,\,260$.}
 \label{fig:semi9}
\end{figure}

If the time-scale falls below a certain value (say $t_{\rm GR} \simeq 1000 $) or
the hyperbolic velocity exceeds $0.001 c$, we first activate the radiation term
by itself.
The terms 1PN, 2PN, 3PN are then included for progressively shorter values
chosen experimentally (say 100, 10, 1).
This procedure is based on the assumption that precession only plays an important
role in detuning the eccentricity growth in the later stages, which has been
tested for some idealized cases of nearly isolated two-body motion.
In order to reach the relativistic regime, it is necessary to achieve
considerable shrinkage by dynamical evolution and/or a large eccentricity.
Unless stellar disruption occurs, gravitational coalescence is adopted if the
distance falls inside three Schwarzschild radii, $6 m_0/c^2$.
Note that the two-body elements $a$ and $e$ needed for decision-making are
evaluated consistently with the corresponding post-Newtonian expansion
(Mora \& Will 2004) in order to avoid spurious effects.

An early $N$-body simulation of relativistic effects in compact systems
(Lee 1993) examined binary formation by gravitational radiation capture
using KS regularization.
More recently, this approach was extended up to 2.5PN for a similar
study of collisional runaway (Kupi, Amaro--Seoane \& Spurzem 2006).
A fully consistent scheme was employed, leading to a shorter merging
time-scale for the central object.

\section{Some results}

The special methods described above have been employed to study centrally
concentrated systems containing one or more massive bodies.
First we display the behaviour of a massive binary which formed as the
result of two merging clusters, each with $N_0 = 1.2 \times 10^5$ equal-mass
particles and a single body with mass ratio $m_0/m_i = (2 N_0)^{1/2}$.
Experience shows that a square root relation for the mass is still
sufficient for such a binary to dominate the central region. 
Here the eccentricity increased steadily until relativistic effects became
important ($V^{\ast} \simeq 16\,{\rm km\,s}^{-1}$).
At present there is no good theoretical explanation for this behaviour.
Likewise, a significant shrinkage of the semi-major axis took place.
Most of this was due to dynamical evolution by sling-shot interactions
until the final stage of rapid energy loss which led to gravitational
coalescence.

The onset of the rather steep decrease in figure~\ref{fig:semi4} is due to
the slightly delayed experimental activation of the relativistic terms, which
were included up to order $1/c^5$ (Aarseth 2003b).
However, the eccentricity increase to a large value ($e = 0.996$) by dynamical
effects shown in figure~\ref{fig:ecc4} would also speed up the shrinkage.
Although various refinements have been added over the years, this early
application of the TTL method demonstrated its usefulness in dealing with a
difficult problem.

Simulations with the wheel-spoke method have been made for point-mass
systems as well as cluster models containing white dwarfs of mass
$1\,M_{\odot}$ and radius $r^{\ast} = 5 \times 10^{-5}$\,au.
With only one BH present, the qualitative behaviour is now quite different.
The evolution of a point-mass model is illustrated in figures~\ref{fig:ecc9}
and \ref{fig:semi9}.
In this model, a rather small value of the half-mass radius was used
($r_{\rm h} \simeq 0.1$\,pc or $c \simeq 5000$) and hence the coalescence
distance is larger than in similar models with $r_{\rm h} \simeq 1$\,pc
which also produced 5--6 such events.
A number of large amplitude excursions in eccentricity can be seen.
This behaviour is the hallmark of Kozai oscillations (Kozai 1962) which in
many cases are induced by the second innermost stable orbit (Aarseth 2007).

The corresponding evolution of semi-major axis in figure~\ref{fig:semi9}
is characterized by several episodes of orbital shrinkage, followed by
coalescence.
Each coalescence is controlled by the minimum pericentre distance,
$a(1-e)$, where the semi-major axis has a tendency to decrease as a result
of dynamical evolution (although reversed near the middle).

\begin{figure}
 \includegraphics{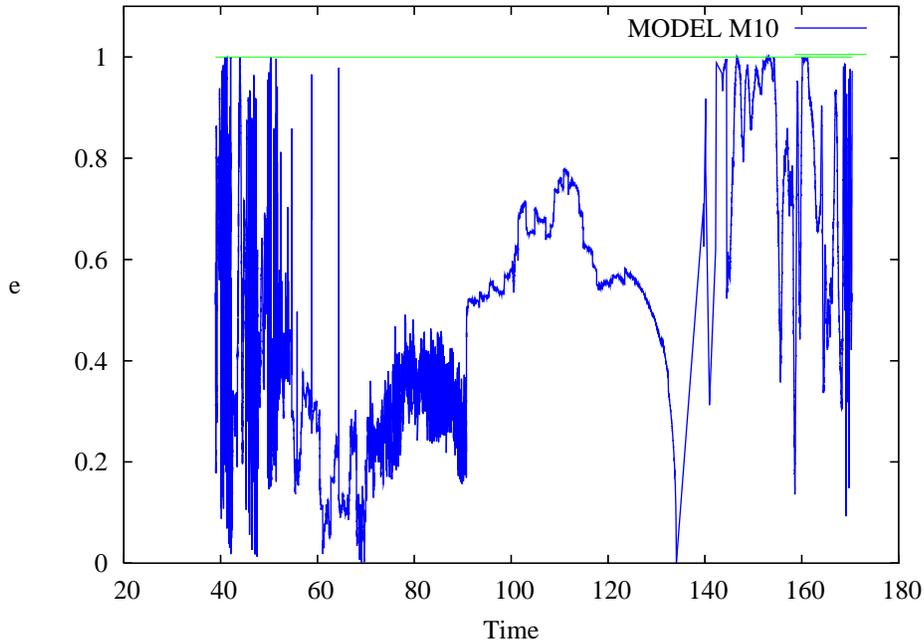}
  \caption{Eccentricity of the innermost binary, model M10.}
 \label{fig:ecc10}
\end{figure}

Several white dwarf models were also investigated using similar initial
conditions (Aarseth 2007).
In such cases, disruption takes place when the pericentre distance falls
below about $7 r^{\ast}$.
Even so, this is sufficiently small for significant energy loss by
gravitational radiation, and hence would act to increase the number of
disruptions which may be observable by LISA technology.

Recent efforts have been concerned with evaluating the algorithmic
regularization code (Mikkola \& Merritt 2006) in an $N$-body environment,
again with the above realistic conditions of a white dwarf population.
Extensive tests with a single massive object confirmed the qualitative
results obtained by the wheel-spoke code.
At the same time, the decision-making for efficient usage was improved in
response to various technical problems.
Very recently, a more general mass distribution was studied.
In order to bypass the generation of appropriate initial conditions, several
simulations were started from $t=39$ of a wheel-spoke model by inserting
artificially two massive objects with eccentric orbits in the inner region.
The code is sufficiently robust to tolerate such a discontinuity and the
evolution progressed normally.

\begin{figure}
 \includegraphics{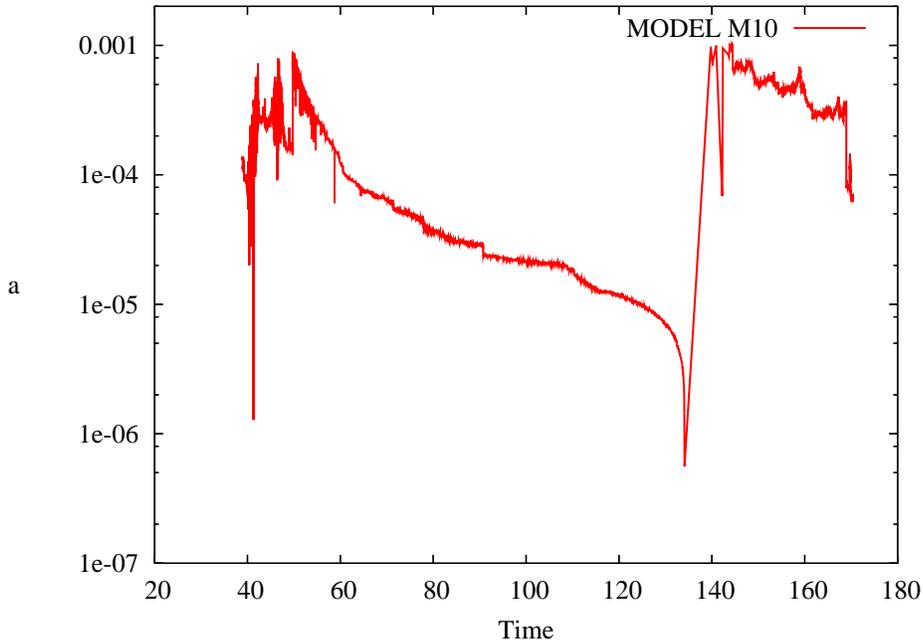}
  \caption{Semi-major axis of the innermost binary, model M10.}
 \label{fig:semi10}
\end{figure}

As expected, the presence of two extra heavy objects leads to orbital
decay towards the centre by dynamical friction.
In the first such attempt (model $M10$), the extra masses were quite modest
($4 \times 10^{-4}$ and $3 \times 10^{-4}$) with a favourable value for the
rms velocity, $V^{\ast} \simeq 300\,\rm km \,s^{-1}$
(or $c = 1000$ in $N$-body units).
Consequently the early evolution in figures~\ref{fig:ecc10} and
\ref{fig:semi10} was qualitatively similar to some previous models, with six
coalescence events due to the Kozai mechanism.
At a later stage, all three heavy objects formed a compact subsystem, after
which the lightest was ejected but not with sufficient energy to escape.
There followed a long period with the dominant binary shrinking slowly by
GR energy loss until coalescence.
The third BH also returned to the centre before this event and underwent
coalescence some 10 time units later.
Consequently, the later stage was again dominated by a single massive body and
three more mergers took place.

\begin{figure}
 \includegraphics{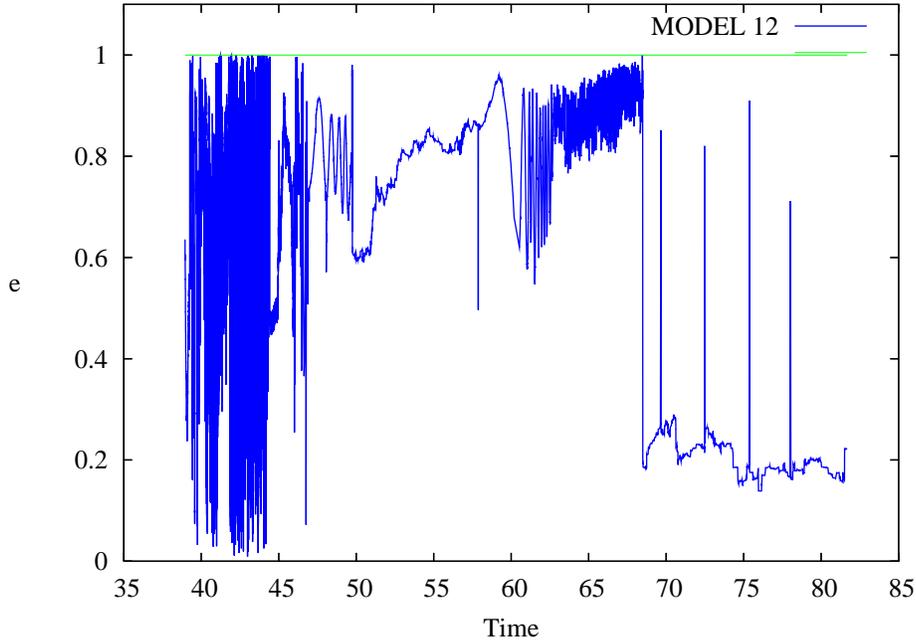}
  \caption{Eccentricity of the innermost binary, model M12.}
 \label{fig:ecc12}
\end{figure}

We also report on two models ($M11$ and $M12$) with more
realistic parameters.
Two additional BHs were added with masses
$2 \times 10^{-3}$ and $1 \times 10^{-3}$ in a less dense system with
$V^{\ast} \simeq 20\,\rm km \,s^{-1}$.
Consequently, $c = 15\,000$ was adopted.
Starting from $t=39$ as above, the second and third BHs already formed part of
the compact subsystem at $t \simeq 44$ and $t \simeq 48$, respectively.
The latter was ejected but returned for an energetic sling-shot interaction
(Saslaw, Valtonen \& Aarseth 1974) leading to escape from the cluster.
The subsequent semi-major axis declined to
$a_{\rm BH} \simeq 3.5 \times 10^{-5}$ at $t = 112$ when the calculation was
halted, with the GR time-scale still quite large ($e \simeq 0.66$).
It is noteworthy that the final binding energy exceeded $33\,\%$ of the total
energy, all due to dynamical effects.

\begin{figure}
 \includegraphics{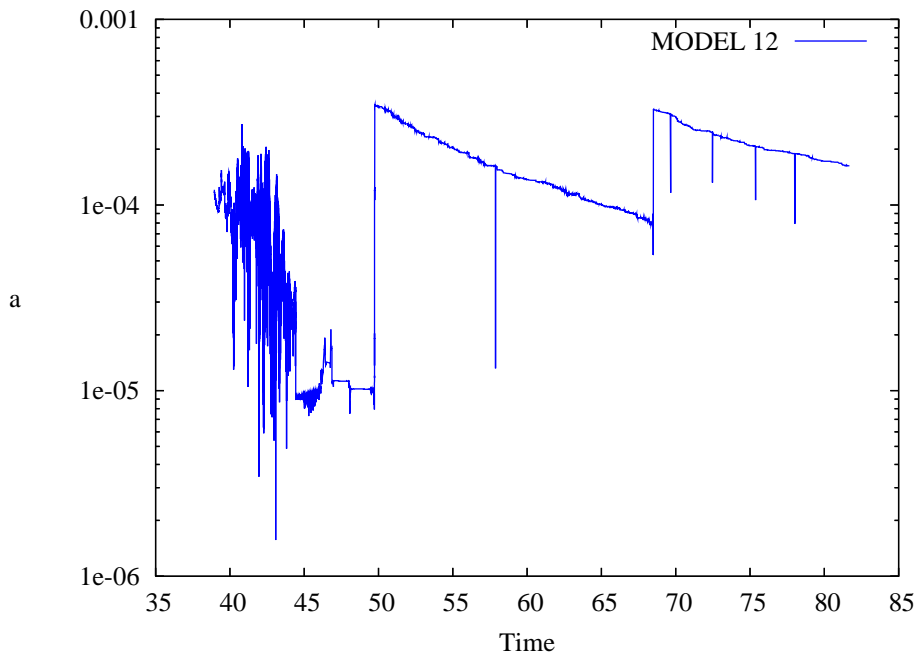}
  \caption{Semi-major axis of the innermost binary, model M12.}
 \label{fig:semi12}
\end{figure}

Model $M12$ exhibited another type of behaviour, illustrated in
figures~\ref{fig:ecc12} and \ref{fig:semi12}.
The early stage ($t < 46$) gave rise to Kozai oscillations up to $e=0.9999$
involving the primary BH but only sufficient to touch the relativistic
regime.
The second BH formed a binary with the first soon after ($t \simeq 50$),
whereupon the innermost binary became wider.
In view of the considerably smaller rms velocity, relativistic effects only
played a minor role.
However, near $t = 68$, the eccentricity of the dominant binary
($M = 5 \times 10^{-3}$) did reach $e_{\rm max} = 0.99998$, induced by the
third BH with small eccentricity and large inclination, which resulted in
GR coalescence.
This model therefore ended up with a low-eccentricity BH binary,
decreasing slowly in size by dynamical means only.

\section{The future}

In this contribution, we have demonstrated the practical usefulness of three
special regularization methods for treating one or more black holes in a
dense stellar system.
Moreover, each method is capable of dealing with considerable dynamical
shrinkage and the calculation can be extended to the relativistic regime if
necessary.
Such problems offer a rich complexity of outcomes for future studies.
In particular, it will be desirable to extend the simulations to larger systems.
Based on present experience, a careful treatment is needed to deal with short
time-scales which are a feature of compact subsystems.
Observational imprints of such interactions will undoubtedly reveal new secrets.

\begin{acknowledgments}
I am greatly indebted to Dr Seppo Mikkola for providing the stand-alone 
algorithmic regularization code as well as much technical advice.
\end{acknowledgments}

\end{document}